%% file: main.tex
\documentclass[pdftex]{llncs}

\input{preamble}
\input{macros}

\title{Oink: an Implementation and Evaluation of Modern Parity Game Solvers}

\author{
  Tom van Dijk\thanks{The author is supported by FWF, NFN Grant S11408-N23 (RiSE)}
}

\institute{
  Formal Models and Verification \\ Johannes Kepler University, Linz \\
  \email{tom.vandijk@jku.at}
}

\begin{document}
\maketitle
\begin{abstract}
\input{abstract}
\end{abstract}

\section{Introduction}
\label{sec:introduction}
\input{intro}

\section{Preliminaries}
\label{sec:preliminaries}
\input{preliminaries}

\section{Oink}
\label{sec:oink}
\input{theoink}

\section{Strategy Improvement}
\label{sec:si}
\input{si}

\section{Progress Measures}
\label{sec:pm}
\input{spm}

\section{Zielonka's Recursive Algorithm}
\label{sec:zielonka}
\input{zielonka}

\section{Priority Promotion}
\label{sec:prioritypromotion}
\input{prioprom}

\section{Conclusions}
\label{sec:conclusions}
\input{conclusions}

\clearpage

\bibliographystyle{splncs03}
\bibliography{lit}

\end{document}

%% file: preamble.tex
\usepackage{etex}
\usepackage[T1]{fontenc}
\usepackage{textcomp}
\usepackage[utf8]{inputenc}

\usepackage{multicol,multirow,colortbl,array,booktabs}
\usepackage{amsmath,amssymb,mathtools}
\usepackage{microtype}

\usepackage{float}
\usepackage{graphicx}
\usepackage{enumerate}
\usepackage{xspace}
\usepackage{listings}
\usepackage{url}
\usepackage[colorlinks,citecolor=blue,linkcolor=red]{hyperref}

\usepackage[british]{babel}
\usepackage{tabularx}

\usepackage{amssymb}
\usepackage{amsmath}
\usepackage{subfig}

\usepackage{tabu}
\usepackage{booktabs}
\usepackage{longtable}
\usepackage[inline]{enumitem}
\usepackage{multicol}

\usepackage{tikz}
\usetikzlibrary{automata, arrows, calc, shapes, shapes.misc, shapes.multipart, decorations.pathreplacing, decorations.markings, decorations.pathmorphing, fit, patterns, positioning}
\usetikzlibrary{shapes.geometric}

\usepackage{comment}
\usepackage{todonotes}
\usepackage{soul}

\usepackage[noline,noend]{algorithm2e}
\AlgoDontDisplayBlockMarkers
\DontPrintSemicolon
\LinesNumbered
\SetAlgoCaptionLayout{normal}
\SetAlgoCaptionSeparator{}
\SetAlCapSkip{.5em}
\makeatletter
\renewcommand{\algocf@makecaption}[2]{%
	\addtolength{\hsize}{1.5\algomargin}%
	\parbox[t]{\hsize}{\algocf@captiontext{#1:}{#2}}%
	\addtolength{\hsize}{-1.5\algomargin}%
}
\makeatother
\SetKwProg{Def}{def}{:}{}
\SetKwProg{Fn}{def}{:}{}
\SetKwProg{Thread}{thread}{:}{}
\SetKwIF{If}{ElseIf}{Else}{if}{:}{elif}{else:}{}
\SetKwFor{For}{for}{:}{}
\SetKwFor{While}{while}{:}{}
\SetKwIF{Catch}{}{Try}{catch}{:}{}{try:}{}
\SetKwFor{Loop}{loop:}{}{}
\SetKwFor{DoPar}{do in parallel:}{}{}
\SetArgSty{textup}
\SetFuncArgSty{textrm}
\SetKw{Break}{break}
\SetKw{Raise}{raise}
\SetKw{False}{false}
\SetKw{True}{true}
\SetKwFunction{Zielonka}{zielonka}
\SetKwFunction{ZielonkaAttr}{attr}
\SetKwFunction{StrategyImprovement}{strategy-improvement}
\SetKwFunction{ComputeValuations}{compute-valuations}
\SetKwFunction{ComputeValuation}{compute-valuation}
\SetKwFunction{BackwardUpdate}{backwards-update}

\usepackage{wrapfig}

\usepackage{tikz}
\usetikzlibrary{arrows, automata, backgrounds, calc, chains, fit, intersections, shadows, patterns, positioning, shapes, shapes.geometric, shapes.misc, trees, decorations.pathreplacing, decorations.markings, decorations.pathmorphing}

\tikzstyle{oval} = [state, ellipse, minimum size=4mm, inner sep=0.5mm, node distance=1cm]
\tikzstyle{leaf}=[rectangle,draw=black!70,solid,thick,minimum size=6mm]
\tikzstyle{input}=[oval,draw=black!70,solid,thick,minimum size=6mm]
\tikzstyle{andgate}=[regular polygon, regular polygon sides=3, draw=black!70,solid,thick,inner sep=0.5mm]
\tikzstyle{latch}=[rectangle,draw=black!70,solid,thick,minimum size=6mm]
\tikzstyle{var}=[oval,draw=black!70,solid,thick,minimum size=6mm]
\tikzstyle{p}=[->,solid,thick,draw]
\tikzstyle{n}=[->,dashed,thick,draw]
\tikzstyle{comp}=[circle, fill=black, draw=black, minimum size=2mm, inner sep=0pt]

\usepackage{xcolor}% http://ctan.org/pkg/xcolor

\makeatletter
\newcommand{\algcolor}[2]{%
  \hskip-\ALG@thistlm\colorbox{#1}{\parbox{\dimexpr\linewidth-2\fboxsep}{\hskip\ALG@thistlm\relax #2}}%
}

\makeatother

%% file: macros.tex
\newcommand{\Attr}{\mathit{Attr}}
\newcommand{\pr}{\textsf{pr}}

\newcommand{\invalpha}{{\overline{\alpha}}}

\newcommand{\sqdiamond}{\tikz [x=1.2ex,y=1.2ex,line width=.08ex] \draw (0,.5) -- (.5,1) -- (1,.5) -- (.5,0) -- (0,.5) -- cycle;}
\newcommand{\sqsq}{\tikz [x=0.95ex,y=1ex,line width=.1ex] \draw (0,0) -- (1,0) -- (1,1) -- (0,1) -- (0,0) -- cycle;}

\newcommand{\Even}{{\scalebox{0.95}{\sqdiamond}}}
\newcommand{\Odd}{{\scalebox{0.9}{$\sqsq$}}}

\newcommand{\Veven}{V_{\Even}}
\newcommand{\Vodd}{V_{\Odd}}

%% file: abstract.tex
%!TEX root = main.tex
Parity games have important practical applications in formal verification and synthesis,
especially to solve the model-checking problem of the modal mu-calculus. They are
also interesting from the theory perspective, as
they are widely believed to admit a polynomial solution, but so far no such algorithm is known.
In recent years, a number of new algorithms and improvements to
existing algorithms have been proposed. We implement a new and easy
to extend tool Oink, which is a high-performance implementation of modern
parity game algorithms.  We further present a comprehensive
empirical evaluation of modern parity game algorithms and solvers, both on real world
benchmarks and randomly generated games.  Our experiments show that
our new tool Oink outperforms the current state-of-the-art.

% \keywords{multi-core, parallel, parity games, strategy improvement, small progress measures, Zielonka, priority promotion}

%% file: intro.tex
%!TEX root = main.tex

Parity games are turn-based games played on a finite graph.
Two players \emph{Odd} and \emph{Even} play an infinite game by moving a token along the edges of the graph.
Each vertex is labeled with a natural number \emph{priority} and the winner of the game is determined by the parity of the highest priority that is encountered infinitely often.
Player Odd wins if this parity is odd; otherwise, player Even wins.

Parity games are interesting both for their practical applications and for complexity theoretic reasons.
Their study has been motivated by their relation to many problems in formal verification and synthesis that can be reduced to the problem of solving parity games, as parity games capture the expressive power of nested least and greatest fixpoint operators~\cite{DBLP:conf/cav/Fearnley17}.
In particular, deciding the winner of a parity game is polynomial-time equivalent to checking non-emptiness of non-deterministic parity tree automata~\cite{DBLP:conf/stoc/KupfermanV98}, and to the explicit model-checking problem of the modal $\mu$-calculus~\cite{DBLP:conf/cav/EmersonJS93,DBLP:journals/tcs/EmersonJS01,DBLP:conf/dagstuhl/2001automata,DBLP:journals/tcs/Kozen83}.

Parity games are interesting in complexity theory, 
as the problem of determining the winner of a parity game is known to lie in $\text{UP}\cap\text{co-UP}$~\cite{DBLP:journals/ipl/Jurdzinski98},
as well as in $\text{NP}\cap\text{co-NP}$~\cite{DBLP:journals/tcs/EmersonJS01}.
This problem is therefore unlikely to be NP-complete and it is widely believed that a polynomial solution may exist.
Despite much effort, no such algorithm has yet been found

Motivated by recent publications with both novel approaches and improvements to known algorithms,
we implement a number of modern solvers in our new tool \textbf{Oink}, which aims to provide a high-performance implementation of parity game solvers.
Oink is designed as a library that integrates with other tools and can easily be extended.
We use Oink to provide a modern empirical evaluation of parity game solvers based on both real world benchmarks and randomly generated games.

In past publications new and improved algorithms are often tested against the implementation of Zielonka's algorithm in the PGSolver tool~\cite{DBLP:conf/atva/FriedmannL09}.
However, various recent publications~\cite{DBLP:conf/iccS/ArcucciMMS17,DBLP:conf/tase/LiuDT14,DBLP:conf/facs2/StasioMPS14} suggest that much better performance can be obtained.
We combine a number of improvements from the literature~\cite{DBLP:conf/tase/LiuDT14,DBLP:conf/facs2/StasioMPS14,Verver2013} and propose additional optimizations.
We show that our implementation of Zielonka's algorithm outperforms PGSolver by several orders of magnitude.

We describe Oink in Sec.~\ref{sec:oink} and provide accessible descriptions of the implemented state-of-the-art algorithms in Sec.~\ref{sec:si}--\ref{sec:prioritypromotion}.
We implement the \emph{strategy improvement} algorithm (Sec.~\ref{sec:si}),
both the \emph{small progress measures} and the recently proposed \emph{quasi-polynomial progress measures} algorithms (Sec.~\ref{sec:pm}),
the well-known \emph{Zielonka} algorithm (Sec.~\ref{sec:zielonka}) as well as
a number of related algorithms from the \emph{priority promotion} family (Sec.~\ref{sec:prioritypromotion}).
We also propose an alternative multi-core implementation of {strategy improvement}.

%% file: preliminaries.tex
%!TEX root = main.tex

Parity games are two-player turn-based infinite-duration games over a finite directed graph $G=(V,E)$, where
every vertex belongs to exactly one of two players called player \emph{Even} and player \emph{Odd},
and where every vertex is assigned a natural number called the \emph{priority}.
Starting from some initial vertex, a play of both players is an infinite path in $G$ where the owner of each vertex determines the next move. The winner of such an infinite play is determined by the parity of the highest priority that occurs infinitely often along the play.

More formally, a parity game $\Game$ is a tuple $(\Veven, \Vodd, E, \pr)$ where $V=\Veven\cup \Vodd$ is a set of vertices partitioned into the sets $\Veven$ controlled by player \emph{Even} and $\Vodd$ controlled by player \emph{Odd}, and $E\subseteq V\times V$ is a total relation describing all possible moves, i.e., every vertex has at least one successor.
We also write $E(u)$ for all successors of $u$ and $u\rightarrow v$ for $v\in E(u)$.
The function $\pr\colon V\rightarrow \{0,1,\dotsc,d\}$
assigns to each vertex a \emph{priority}, where $d$ is the highest priority in the game.

We write $\pr(v)$ for the priority of a vertex $v$
and $\pr(V)$ for the highest priority of a set of vertices $V$ and
$\pr(\Game)$ for the highest priority in the game $\Game$.
Furthermore, we write $\pr^{-1}(i)$ for all vertices with the priority $i$.
A \emph{path} $\pi=v_0 v_1 \dots$ is a sequence of vertices consistent with $E$, i.e.,
$v_i \rightarrow v_{i+1}$ for all successive vertices.
A \emph{play} is an infinite path.
We write $\pr(\pi)$ for the highest priority in $\pi$ is $\pi$ is finite, or the highest priority that occurs infinitely often if $\pi$ is infinite.
Player Even wins a play $\pi$ if $\pr(\pi)$ is even; player Odd wins if $\pr(\pi)$ is odd.

A \emph{strategy} $\sigma\colon V\to V$ is a partial function that assigns to each vertex in its domain a single successor in $E$, i.e., $\sigma\subseteq E$.
We typically refer to \emph{a strategy of player $\alpha$} to restrict $\sigma$ to all vertices controlled by player $\alpha$.
A player wins a vertex if they have a strategy such that all plays consistent with this strategy are winning for the player.
A fundamental result for parity games is that they are memoryless determined~\cite{DBLP:conf/focs/EmersonJ91}, i.e., each vertex is winning for exactly one player, and both players have a strategy for each of their winning vertices.

Algorithms for solving parity games frequently employ (variations of) \emph{attractor computation}.
Given a set of vertices $A$,
the attractor of $A$ for a player $\alpha$ represents those vertices that $\alpha$ can force the play toward.
We write $\Attr^\Game_\alpha(A)$ to attract vertices in $\Game$ to $A$ as player $\alpha$,
i.e.,
\[\mu Z \,.\, A \cup \{\;v\in V_\alpha \mid E(v)\cap Z \neq \emptyset\;\} \cup \{\;v\in V_{\invalpha} \mid E(v)\subseteq Z\;\}\]
Informally, we compute the $\alpha$-attractor of $A$ by iteratively adding vertices to $A$ of $\alpha$ that have a successor in $A$ and of $\invalpha$ that have no successors outside $A$.

\subsection{Solvers}

We briefly introduce several approaches to solving parity games.
These approaches can be roughly divided into two categories.

\emph{First}, several algorithms iteratively perform local updates to vertices
until a fixed point is reached. Each vertex is equipped with some measure
which records the best game either player knows that they can play from that vertex so far.
By updating measures based on the successors, they play the game backwards.
The final measures indicate the
winning player of each vertex and typically a winning strategy for one or both players.
The \emph{strategy improvement} algorithm (Sec.~\ref{sec:si}) and the \emph{progress measures} algorithms (Sec.~\ref{sec:pm}) fall into this category.

\emph{Second}, several algorithms employ attractor computation to partition the game into regions that share a certain
property.
This partition is refined until the winners of some or all vertices can be identified, as well as the strategy for the winning player(s).
The \emph{recursive Zielonka} algorithm (Sec.~\ref{sec:zielonka}) and the recently proposed \emph{priority promotion} algorithms (Sec.~\ref{sec:prioritypromotion})
fall into this category.

\subsection{Empirical evaluation}
\label{sec:design}

Our goal in the empirical study is three-fold.
\emph{First}, we aim to compare modern algorithms and treat them fairly.
We therefore need to establish that our implementation is competitive with existing work.
\emph{Second}, we compare the algorithms that are implemented in Oink directly.
\emph{Third}, as two algorithms have a parallel implementation, we also study the obtained parallel speedup and the parallel overhead when going from a sequential to a multi-core implementation.

We use the parity game benchmarks from model checking and equivalence checking proposed by Keiren~\cite{DBLP:conf/fsen/Keiren15} that are publicly available online.
This is a total of $313$ model checking and $216$ equivalence checking games.
We also consider different classes of random games,
in part because the literature on parity games tends to favor studying the behavior of algorithms on random games.
We include three classes of self-loop-free random games generated using PGSolver with a fixed number of vertices:
\begin{itemize}
	\item low out-degree random games (\texttt{randomgame N N 1 2 x}) \\
	$N \in \{\, 100, 200, 500, 1000, 2000, 5000, 10000, 20000 \,\}$
	\item fully random games (\texttt{randomgame N N 1 N x}) \\
	$N \in \{\, 100, 500, 1000, 2000, 4000 \,\}$
	\item low-degree steady games (\texttt{steadygame N 1 4 1 4})\\
	$N \in \{\, 100, 200, 500, 1000, 2000, 5000, 10000, 20000 \,\}$
\end{itemize}
We generate $20$ games for each parameter $N$, in total $420$ random games.
We include low-degree games,
since the solvers may behave differently on games where all vertices have few edges.

We present the evaluation in two ways.
We compare runtimes of algorithms and penalize algorithms that do not finish on time (with a timeout of $15$ minutes) by a factor $2\times$ (PAR2), i.e., we assume that their runtime is $2\times$ the timeout.
This may still be quite optimistic. % for algorithms with many timeouts.
Compared to a timeout of $10$ minutes, only few more games could be solved in $15$ minutes. %Increasing the timeout from $10$ minutes to $15$ minutes only resulted in few more solved games.
We also generate so-called \emph{cactus plots} (often used to compare solvers in the SAT community) that show that a solver solved $X$ models within $Y$ seconds individually.

All experimental scripts and log files are available online via \url{http://www.github.com/trolando/oink-experiments}.
The experiments were performed on a cluster of Dell PowerEdge M610 servers with two Xeon E5520 processors and 24GB internal memory each.
The tools were compiled with gcc 5.4.0.

%% file: theoink.tex
%!TEX root = main.tex

We study modern parity game algorithms using our research tool named \textbf{Oink}.
Oink is written in C++ and is publicly available under a permissive license via \url{https://www.github.com/trolando/oink}.
Oink is easy to extend, as new solvers subclass the \verb|Solver| class and only require a few extra lines in \verb|solvers.cpp|.

Apart from implementing the full solvers described below,
Oink also implements several preprocessors similar to other parity game solvers.
We base our choices mainly on the practical considerations and observations by Friedmann and Lange~\cite{DBLP:conf/atva/FriedmannL09} and by Verver~\cite{Verver2013}.
We always reorder the vertices by priority and renumber the priorities from $0$ to eliminate gaps (not the same as compression).
The former is beneficial for the attractor-based algorithms in Sec.~\ref{sec:zielonka} and~\ref{sec:prioritypromotion}.
The latter may reduce the amount of memory required for the measures-based algorithms in Sec.~\ref{sec:si} and~\ref{sec:pm}.

The following preprocessors are optional.
Oink can perform priority inflation and priority compression, as described in~\cite{DBLP:conf/atva/FriedmannL09}.
% We do not implement priority propagation, as this typically does not result in faster computation~\cite{DBLP:conf/atva/FriedmannL09}.
We implement self-loop solving and winner-controlled winning cycle detection, as proposed in~\cite{Verver2013}.
Winner-controlled winning cycle detection is a prerequisite for the strategy improvement algorithm of Sec.~\ref{sec:si} but is optional for the other algorithms.
We trivially solve games with only a single parity.
Finally, we also implement SCC decomposition, which repeatedly solves a bottom SCC of the game until the full game is solved.

The correctness of an algorithm does not imply that implementations are correct.
Although a formal proof of the implementations would be preferred,
we also implement a fast solution verifier
and verify all obtained solutions.

%% file: si.tex
%!TEX root = main.tex

Strategy improvement is a technique where each player iteratively improves their strategies until they are optimal.
Strategy improvement algorithms were first explored for parity games by Jurdziński and Vöge~\cite{DBLP:conf/cav/VogeJ00} and have been subsequently improved in \cite{DBLP:journals/dam/BjorklundV07,DBLP:conf/lpar/Fearnley10,DBLP:conf/gandalf/Friedmann2010,DBLP:journals/corr/abs-0806-2923,DBLP:conf/csl/Schewe08}.
Recently, parallel implementations have been studied for the GPU~\cite{DBLP:conf/cav/Fearnley17,DBLP:conf/atva/HoffmannL13,DBLP:conf/atva/MeyerL16}.
Fearnley~\cite{DBLP:conf/cav/Fearnley17} also implements their parallel algorithm for multi-core CPUs.
Our treatment in this section is mostly based on~\cite{DBLP:conf/cav/Fearnley17,DBLP:journals/corr/abs-0806-2923}.

In the strategy improvement algorithm, player Even has a strategy $\sigma$
and player Odd has a strategy $\tau$ for all their vertices.
They improve their strategies until a fixed point is reached,
at which point the game is solved.
Instead of choosing a successor, player Even may also end the play.
In the specific algorithm here, player Even delays selecting a strategy for a vertex until there is a favorable continuation. 
As $\sigma$ and $\tau$ cover all vertices, they induce a fixed path from each vertex.
This path is either infinite (a play) or ends at a vertex of player Even.
The strategy is evaluated by computing a valuation for each vertex based on the current paths.
Strategies are improved by choosing the most favorable successor.

The valuation used in e.g.~\cite{DBLP:conf/cav/Fearnley17,DBLP:journals/corr/abs-0806-2923}
assigns to (infinite) plays the value $\top$ and to (finite) paths a function $L(p)$ that records for each priority $p$ how often it occurs in the path.
To determine the best move for each player,
a total order $\sqsubset$ compares valuations as follows.
For non-$\top$ valuations $L_1$ and $L_2$, $L_1 \sqsubset L_2$ iff there exists a highest priority $z$ that is different in $L_1$ and $L_2$, i.e., $z = \max \,\{\, z \mid L_1(z) \neq L_2(z) \,\}$, and either $L_1(z)<L_2(z)$ if $z$ is even, or $L_1(z)>L_2(z)$ if $z$ is odd.
Furthermore, $L\sqsubset\top$ for any $L\neq\top$.
If $L_1 \sqsubset L_2$, then $L_2$ is more favorable for player Even and $L_1$ is more favorable for player Odd.

Intuitively, player Even likes plays where higher even priorities occur more often.
Furthermore, player Even will end the play unless the highest priority in the continuation is even.
Thus infinite paths are won by player Even and the valuation $\top$ represents this.
Player Even will always play to $\top$ and player Odd will always try to avoid $\top$.
This assumes that no winner-controlled winning cycles exist where player Odd wins, which can be guaranteed using a preprocessing step that removes these cycles.

For every strategy $\sigma$ of player Even, a so-called \emph{best response} $\tau$ of player Odd minimizes the valuation of each position.
Player Even always plays against this best response.
In each iteration, after player Odd computes their best response,
player Even computes all \emph{switchable edges} based on the current valuation $L$ and the current strategy $\sigma$.
An edge $(u,v)$ is \emph{switchable} if $L_v \sqsupset L_{\sigma(u)}$.
Not all switchable edges need to be selected for the improved strategy, but as argued in~\cite{DBLP:conf/cav/Fearnley17}, the \emph{greedy all-switches} rule that simply selects the best edge (maximal in $\sqsubset$) for every position performs well in practice.

There are different methods to compute the best response of player Odd.
We refer again to~\cite{DBLP:conf/cav/Fearnley17} for a more in-depth discussion.
Player Odd can compute their best response by repeatedly switching 
all \emph{Odd-switchable} edges, i.e., edges $(u,v)$ s.t. $L_v \sqsubset L_{\tau(u)}$ and $L_v$ is minimal in $\sqsubset$ of all successors of $u$.

\begin{algorithm}[t]
	\Def{\StrategyImprovement}{
		$\sigma$ $\leftarrow$ $(V_0 \mapsto \bot)$, $\tau$ $\leftarrow$ random strategy for Odd \;
		\Repeat{$S_\text{Even}=\emptyset$}{
			\Repeat{$S_\text{Odd}=\emptyset$}{
				\ComputeValuations{$V$, $\sigma \cup \tau$, $L$} \; 
				$\tau$ $\leftarrow$ $\tau[S_\text{Odd}]$ where $S_\text{Odd} = \text{All}_\text{Odd}(\Game,\tau,L)$ \;
			}
		    \texttt{mark-won}($\{v\in V\colon L_v = \top\}$) \;
			$\sigma$ $\leftarrow$ $\sigma[S_\text{Even}]$ where $S_\text{Even} = \text{All}_\text{Even}(\Game,\sigma,L)$ \;
		}
		\Return $(W_0, W_1, \sigma, \tau)$ where $W_0$ $\leftarrow$ $\{v\in V\colon L_v = \top\}$, $W_1$ $\leftarrow$ $V \setminus W_0$ 
	}
\caption{The strategy improvement algorithm.}
\label{alg:strimpr}
\end{algorithm}

The players thus improve their strategies as in Algorithm~\ref{alg:strimpr},
where $\text{All}_\text{Odd}$ and $\text{All}_\text{Even}$ compute all switchable edges as described above.
The initial strategy for player Even is to always end the play.
Player Odd computes their best response, starting from a random $\tau$ initially.
Player Even then improves their strategy once.
They improve their strategies until a fixed point is reached.
Then all vertices with valuation $\top$ are won by player Even with strategy $\sigma$ and all other vertices are won by player Odd with strategy $\tau$~\cite{DBLP:conf/cav/Fearnley17}.
We extend the algorithm given in~\cite{DBLP:conf/cav/Fearnley17} at line~8 by marking vertices with valuation $\top$ after player Odd computes their best response as ``won''. We no longer need to consider them for $\text{All}_\text{Odd}$ and $\text{All}_\text{Even}$ as player Odd was unable to break the infinite play and thus they are won by Even.

\begin{algorithm}[t]
	\Def{\BackwardUpdate{$v$, \texttt{into}, $L$}}{
		\For{$u \in \texttt{into}(v)$}{
			$L_u$ $\leftarrow$ $L_{v} [ \pr(u)\mapsto L_{v}(\pr(u))+1  ]$ \;
			\textbf{spawn} \ComputeValuation{$u$, $\sigma$, $L$}
		}
		\textbf{sync all} \;
	}
	\BlankLine
	\Def{\ComputeValuations{$V$, $\sigma$, $L$}}{
		\textbf{parallel} \lFor{$v\in V$}{$L_v$ $\leftarrow$ $\top$ ; \texttt{into}($v$) $\leftarrow$ $\emptyset$}
		\textbf{parallel} \lFor{$v\in V \mid \sigma(v)\neq\bot$}{add $v$ to \texttt{into}($\sigma(v))$}
		\textbf{parallel} \For{$v\in V \mid \sigma(v)=\bot$}{
			$L_v$ $\leftarrow$ $0 \, [ \pr(v)\mapsto 1 ]$ \;
			\BackwardUpdate{$v$, \texttt{into}, $L$}
		}
	}
	\caption{Computing valuations in parallel.}
	\label{alg:computevals}
\end{algorithm}

The valuations can be computed in different ways.
Fearnley implements a parallel algorithm that uses list ranking in two steps. The first step computes an Euler tour of the game restricted to chosen strategies $\sigma$ and $\tau$ resulting in a list. The second step uses a three-step parallel reduction algorithm to sum all values of the list. The list is divided into sublists which are each summed independently in the first sweep, all subresults are then propagated in the second sweep and then the final values are computed in the third sweep. See further~\cite{DBLP:conf/cav/Fearnley17}.

We propose an alternative parallel algorithm to compute the valuation.
We start from each Even vertex where the path ends and perform updates along a recursive backwards search, processing predecessors in parallel using task parallelism.
Any vertex that is not visited has valuation $\top$.
See Algorithm~\ref{alg:computevals}.
This algorithm is implemented in Oink using the high-performance work-stealing framework Lace~\cite{DBLP:conf/europar/DijkP14}.
When updating the valuations in $L$, we first sweep twice over all vertices to initialize $L$ for each vertex to $\top$ and to add all vertices to $\texttt{into}(v)$ that have their strategy to $v$.
We also implement computing switchable edges in parallel
via a straight-forward binary reduction using Lace.

\subsubsection{Empirical evaluation}

\begin{table}[t]
	\begin{tabu} to \linewidth {X[2]X[2r]X[c]X[2r]X[c]X[2r]X[c]X[2r]X[c]}
		\toprule
		& \multicolumn{2}{c}{Model checking} & \multicolumn{2}{c}{Equiv checking} & \multicolumn{2}{c}{Random games} & \multicolumn{2}{c}{Total} \\
		\midrule
		\input{R/psi}
		\bottomrule
	\end{tabu}
	\vspace{.5em}
	\caption{Runtimes in sec. (PAR2) and number of timeouts (15 minutes) of the three solvers PGSolver (\texttt{pgsi}), the solver by Fearnley~\cite{DBLP:conf/cav/Fearnley17} with sequential (\texttt{parsi-seq}) and multi-core variants, and Oink with sequential (\texttt{psi}) and multi-core variants.}
	\label{tbl:psi}
\end{table}

We compare the performance of Oink with the sequential and parallel solvers ($1$ or $8$ threads) by Fearnley~\cite{DBLP:conf/cav/Fearnley17} and the ``optstratimprov'' solver in PGSolver. We disable optional preprocessing in all solvers.
We only consider games without winner-controlled winning cycles, which are $289$ model checking, $182$ equivalence checking and $279$ random games, in total $750$ games.

See Table~\ref{tbl:psi}.
We observe that PGSolver is vastly outperformed by Oink and the sequential solver of Fearnley. PGSolver timed out for $160$ games, whereas \texttt{psi} and \texttt{parsi-seq} only timed out for $1$ and $5$ models, respectively.
We observe similar parallel speedup for the parallel solvers, although Fearnley's solver has more overhead from sequential to parallel with $1$ thread.
This might be due to the extra work to produce the Euler tour and to perform list ranking.
The speedup we obtain with Oink is not very impressive, but the vast majority of the games are solved within $1$ second already.
Furthermore, \texttt{psi} and \texttt{parsi-seq} are fairly close in performance.
This is not a surprise, as their implementations are similar; the main difference is that Fearnley uses a forward search and we use a backward search.
Hence, Oink is faster, but not by a large margin.
Finally, we remark that Fearnley reports excellent results for list ranking on GPUs, whereas our algorithm is designed for a multi-core architecture.

%% file: R/psi.tex
% latex table generated in R 3.4.3 by xtable 1.8-2 package
% 
  psi-8 & \textbf{694} & 0 & \textbf{1078} & 0 & \textbf{315} & 0 & \textbf{2087} & 0 \\ 
  psi & 860 & 0 & 3262 & 0 & 480 & 0 & 4603 & 0 \\ 
  psi-1 & 1190 & 0 & 4090 & 0 & 487 & 0 & 5767 & 0 \\ 
  parsi-seq & 1471 & 0 & 4199 & 0 & 1534 & 0 & 7204 & 0 \\ 
  parsi-8 & 2501 & 1 & 2908 & 0 & 56529 & 27 & 61938 & 28 \\ 
  parsi-1 & 4200 & 1 & 13867 & 6 & 71280 & 39 & 89347 & 46 \\ 
  pgsi & 167596 & 88 & 95407 & 49 & 58839 & 27 & 321842 & 164 \\ 
  

%% file: spm.tex
%!TEX root = main.tex

Progress measures is a technique that assigns to each vertex a monotonically increasing \emph{measure}.
The measure of each vertex is \emph{lifted} based on the measures of its successors.
By lifting vertices, players Even and Odd essentially play the game backwards.
The measure represents
a statistic of the most optimal play so far from the vertex, without storing the plays explicitly.

While progress measures have been used elsewhere, they were introduced for
parity games by Jurdziński~\cite{DBLP:conf/stacs/Jurdzinski00}.
Several improvements to the original algorithm are due to Verver~\cite{Verver2013} and Gazda and Willemse~\cite{DBLP:journals/corr/GazdaW15}.
A number of parallel implementations have been proposed for the Playstation~3~\cite{vanderBerg2010}, for multi-core architectures~\cite{DBLP:conf/hvc/HuthKP11,DBLP:journals/entcs/PolW08} and for GPUs~\cite{BootsmaW13,DBLP:conf/atva/HoffmannL13}.
Furthermore, Chatterjee et al. proposed an implementation using BDDs~\cite{DBLP:conf/csl/ChatterjeeDHL17}.
Different types of progress measures were introduced after the recent breakthrough of a quasi-polynomial time algorithm due to Calude et al.~\cite{DBLP:conf/stoc/CaludeJKL017}, which resulted in the progress measures algorithms by Jurdziński et al.~\cite{DBLP:conf/lics/JurdzinskiL17} and by Fearnley et al.~\cite{DBLP:conf/spin/FearnleyJS0W17}.
This section studies small progress measures~\cite{DBLP:conf/stacs/Jurdzinski00} and quasi-polynomial progress measures~\cite{DBLP:conf/spin/FearnleyJS0W17}.

\subsection{Small progress measures}

The original small progress measures algorithm is due to Jurdziński~\cite{DBLP:conf/stacs/Jurdzinski00}.
We rely on the operational interpretation by Gazda and Willemse~\cite{DBLP:journals/corr/GazdaW15} and propose
the \emph{cap-and-carryover} mechanism to further understand the algorithm.

Progress measures record how favorable the game is for one of the players.
W.l.o.g. we assume \emph{even} progress measures.
Given the highest priority $d$, define $\mathbb{M}^\diamond \subseteq \mathbb{N}^d\cup\{\top\}$ to be the largest set containing $\top$ ($\top \notin \mathbb{N}^d$) and only those $d$-tuples with $0$ (denoted as $\_$) on \emph{odd} positions.
An even progress measure $m\in\mathbb{N}^d$ essentially records for a vertex $v$ how often each even priority $p$ is encountered along the most optimal play (starting at $v$) so far, until a higher priority is encountered, i.e., until $p$ no longer dominates.
Such a prefix of the play is called a $p$-dominated stretch.
Suppose that the sequence of priorities for a given play $\pi$ is \underline{0}\underline{0}10\underline{2}1\underline{2}0\underline{2}321\underline{4}2\underline{6}5\underline{6}201, then $m=\{\, 2 \_ 3 \_ 1 \_ 2 \,\}$, since the play starts with a $0$-dominated stretch containing two $0$s, with a $2$-dominated stretch containing three $2$s, with a $4$-dominated stretch containing one $4$, and with a $6$-dominated stretch containing two $6$s.
Furthermore, the special measure $\top$ represents that the vertex can be won by player Even.

A total order $\sqsubset$ compares measures as follows.
For non-$\top$ measures $m_1$ and $m_2$, $m_1 \sqsubset m_2$ iff there exists a highest priority $z = \max \,\{\, z \mid m_1(z) \neq m_2(z) \,\}$ and $m_1(z)<m_2(z)$.
Furthermore, $m\sqsubset\top$ for all $m\neq\top$.
We define a derived ordering $\sqsubset_p$ by restricting $z$ to priorities $\geq p$.
Examples:
\begin{center}
\begin{tabu} to 0.7\linewidth {X[8r]X[l]X[8l]}
$\{ 1 \_ 1 \_ 1 \}$ & $\sqsubset_0$ & $\{ 0 \_ 0 \_ 2 \}$ \\
$\{ 3 \_ 2 \_ 1 \}$ & $\sqsubset_0$ & $\{ 0 \_ 3 \_ 1 \}$ \\
$\{ 1 \_ 2 \_ 1 \}$ & $\sqsubseteq_1$ & $\{ 0 \_ 2 \_ 1 \}$ \\
$\{ 3 \_ 3 \_ 1 \}$ & $\sqsubseteq_4$ & $\{ 0 \_ 0 \_ 1 \}$ \\
\end{tabu}
\end{center}

To compute the progress measure for vertex $v$ when playing to vertex $w$,
given current measures $\rho\colon V\to \mathbb{M}^\diamond$,
we define $\text{Prog}(\rho, v, w)$ as follows:
$$\text{Prog}(\rho, v, w) := 
\begin{cases}
\min \{\, m \in \mathbb{M}^{\diamond} \mid m \sqsupset_{\pr(v)} \rho(w) \,\} & \quad \pr(v)\text{ is even} \\
\min \{\, m \in \mathbb{M}^{\diamond} \mid m \sqsupseteq_{\pr(v)} \rho(w) \,\} & \quad \pr(v)\text{ is odd} \\
\end{cases}
$$

Prog computes the measure of the play obtained by playing from $v$ to the play recorded in $\rho(w)$.
By choosing the lowest measure $m$ according to $\sqsupseteq_{\pr(v)}$, we ensure that all $m(p)$ for $p <\pr(v)$ are set to $0$.
The inequality is strict for even priorities $\pr(v)$ to ensure that $m(\pr(v))$ increases.

Player Even wants to achieve the highest measure, whereas player Odd wants to achieve the lowest measure.
We define $\text{Lift}(\rho, v)$ as follows:
$$\text{Lift}(\rho, v) = 
\begin{cases}
\rho \,[\,{v \mapsto \max \{\, \rho(v), \max \{\, \text{Prog}(\rho, v, w) \;|\; v \rightarrow w \,\} \,\} }\,] & \quad \text{if } v \in \Veven \\
\rho \,[\,{v \mapsto \max \{\, \rho(v), \min \{\, \text{Prog}(\rho, v, w) \;|\; v \rightarrow w \,\} \,\} }\,] & \quad \text{if } v \in \Vodd \\
\end{cases}
$$

By definition, the Lift operation increases measures monotonically.
For the specific algorithm described here, we also observe that $\text{Prog}(\rho,v,w)\sqsupseteq_{\pr(v)}\rho(w)$ and therefore Lift would even monotonically increase
$\rho$ without taking the maximum of the current measure and the best updated successor measure in a lifting procedure that starts with $\rho=V\mapsto 0$.

If we iteratively lift vertices from $\rho=V\mapsto 0$ using $\mathrm{Lift}$,
eventually some vertex may have a measure $m$ such that $m(p)$ for some $p$ is higher than the number of vertices with priority $p$, i.e., $m(p) > \left\lvert V_p \right\rvert$.
In this case, we know that $m$ represents a play that visits at least one vertex with priority $p$ twice and thus contains a cycle dominated by $p$.
Furthermore, player Odd cannot escape from this cycle unless by playing to a higher losing priority.
This follows from the fact that if player Odd could escape from the cycle, then it would not lift to this measure.
The option to play to the higher losing priority is not considered because a measure to a higher priority is $\sqsupset$ a measure that records a cycle.

We need a mechanism to let player Odd play to the next higher priority if it is forced into a cycle.
However, we cannot let just any vertex play to a higher priority when its measure records a cycle, since some other vertex may escape to a lower higher priority.
Therefore we need a mechanism that finds the lowest escape for player Odd.
Small progress measures achieves this using a \emph{cap-and-carryover} mechanism.
$\mathbb{M}^\diamond$ is restricted such that values for each even priority $p$ may not be higher than $\left\lvert V_p \right\rvert$.
When this cap is reached, $\text{Prog}$ will naturally find a next higher $m$ by increasing the value of higher priorities and eventually reach $\top$.
For example, if we have two vertices of priority 2 and two vertices of priority 4 in a game and there is a self-loop of priority 2, measures increase as follows: $\{ 0 \_ 2 \_ 0 \}$, $\{ 0 \_ 0 \_ 1 \}$, $\{ 0 \_ 1 \_ 1 \}$, $\{ 0 \_ 2 \_ 1 \}$, $\{ 0 \_ 0 \_ 2 \}$, $\{ 0 \_ 1 \_ 2 \}$, $\{ 0 \_ 2 \_ 2 \}$, $\top$.

Thus all vertices involved in a cycle will find their measures slowly rising until the measure of some vertex controlled by Odd becomes equal to the measure when playing to a vertex that is not rising.
This is the lowest escape.
If no such escape is found, then the measures rise until $\top$ and these vertices are won by player Even.
The slowly increasing measures no longer follow the operational interpretation described above, but can be understood as player Odd looking for the lowest escape.

We refer to~\cite{DBLP:conf/stacs/Jurdzinski00} for the proof that the fixed point of applying the above lifting operation solves the parity game, such that vertices with measure $\top$ are won by player Even and all other vertices are won by player Odd with a strategy that chooses the successor for which $\text{Prog}(\rho, v, w)$ is the lowest.

We implement three known improvements.
Improvements~2 and~3 are also implemented by PGSolver~\cite{DBLP:conf/atva/FriedmannL09}.
\begin{enumerate}
\item When a vertex with some even priority $p$ is raised to $\top$, the cap of $p$ may be lowered. The reason is that if a play records priority $p$ $\left\lvert V_p \right\rvert$ times, it either contains a vertex now won by player Even or a cycle of priority $p$~\cite{Verver2013}.
\item Small progress measures only computes the strategy for player Odd according to measures for player Even. We compute both \emph{even} and \emph{odd} measures simultaneously to compute the strategy for both players.
\item In addition, we occasionally halt the lifting procedure to perform an attractor computation for player Even to the set of \emph{even}-liftable vertices. Any vertices not in this set are won by player Odd. We can immediately lift these vertices to $\top$ in the \emph{odd} measures. We perform this analysis also for \emph{odd}-liftable measures to potentially lift vertices to $\top$ in the \emph{even} measures.
\end{enumerate}

\subsection{Quasi-polynomial progress measures}
\label{sec:qpt}

Different types of progress measures were introduced after the recent breakthrough of a quasi-polynomial time algorithm due to Calude et al.~\cite{DBLP:conf/stoc/CaludeJKL017}, which resulted in the progress measures algorithms by Jurdziński et al.~\cite{DBLP:conf/lics/JurdzinskiL17} and by Fearnley et al.~\cite{DBLP:conf/spin/FearnleyJS0W17}.
We only briefly and informally describe the idea of~\cite{DBLP:conf/spin/FearnleyJS0W17}.
(Even) measures are $k$-tuples $M\colon (\mathbb{N}\cup\{\bot\})^k\cup\{\top\}$, which record that the optimal play consists of consecutive stretches that are dominated by vertices with even priority.
For example, in the path $1\underline{2}131\underline{4}23\underline{2}15\underline{6}3\underline{2}1\underline{2}$, all vertices are dominated by each pair of underlined vertices of even priority.
$k$ is such that there are fewer than $2^k$ vertices with even priority in the game.
An $8$-tuple $\{\, 2 \; 2\; 4\; \bot\; 5\; \bot\; 6\; \bot \,\}$ denotes a game with consecutive stretches of $1$, $2$, $4$, $16$ and $64$ even vertices, where the first dominating vertex has priority $M(i)$ and may actually be \emph{odd} instead of even.
If the first dominating vertex has an {odd} priority, then player Even must reach a higher priority before continuing to build a play where they have more dominating even vertices than are in the game.
If player Even can visit more dominating even vertices than are in the game, then at least one of these is visited twice and therefore player Even knows that they can win and lifts to $\top$.

\subsection{Empirical evaluation}

\begin{table}[t]
	\begin{tabu} to \linewidth {X[2]X[2r]X[c]X[2r]X[c]X[2r]X[c]X[2r]X[c]}
	\toprule
	& \multicolumn{2}{c}{Model checking} & \multicolumn{2}{c}{Equiv checking} & \multicolumn{2}{c}{Random games} & \multicolumn{2}{c}{Total} \\
	\midrule
	\input{R/pm}
	\bottomrule
\end{tabu}
\vspace{.5em}
\caption{Runtimes in sec. (PAR2) and number of timeouts (15 minutes) of PGSolver (\texttt{pgspm}), pbespgsolve (\texttt{pbesspm}) and the implementations \texttt{spm} and \texttt{qpt} in Oink.}
	\label{tbl:pm}
\end{table}

We compare our implemention of small progress measures and quasi-polynomial progress measures
to the small progress measures implementation of pbespgsolve that comes with the mCRL2 model checker~\cite{DBLP:conf/tacas/CranenGKSVWW13,Verver2013} and the
implementation of small progress measures in PGSolver~\cite{DBLP:conf/atva/FriedmannL09}.
Unfortunately, the solver used in~\cite{DBLP:conf/spin/FearnleyJS0W17} contains proprietary source code and cannot be compiled and compared.
For this comparison, we disabled optional preprocessing, i.e., removing self-loops, winner-controlled winning cycles and solving single-parity games.

See Table~\ref{tbl:pm}.
Although Fearnley et al.~\cite{DBLP:conf/spin/FearnleyJS0W17} say that the QPT solver is mainly interesting for the theoretical result rather than practical performance, we observe that \texttt{qpt} outperforms the other solvers for random games.
Oink is faster than PGSolver, especially for model checking and equivalence checking.

%% file: R/pm.tex
% latex table generated in R 3.4.3 by xtable 1.8-2 package
% 
  spm & \textbf{3637} & 1 & \textbf{7035} & 0 & 168271 & 93 & \textbf{178944} & 94 \\ 
  qpt & 122549 & 64 & 65310 & 31 & \textbf{66303} & 35 & 254162 & 130 \\ 
  pbesspm & 38397 & 20 & 52422 & 27 & 183742 & 101 & 274561 & 148 \\ 
  pgspm & 88800 & 45 & 59885 & 30 & 320666 & 171 & 469351 & 246 \\ 
  

%% file: zielonka.tex
%!TEX root = main.tex

The algorithm by Zielonka~\cite{DBLP:journals/tcs/Zielonka98} is a recursive solver that despite its relatively bad theoretical complexity is known to outperform other algorithms in practice~\cite{DBLP:conf/atva/FriedmannL09}.
Furthermore, tight bounds are known for various classes of games~\cite{DBLP:journals/corr/GazdaW13}.

Zielonka's recursive algorithm is based on attractor computation.
At each step, given current subgame $\Game$, the algorithm removes the attractor $A := \text{Attr}^\Game_\alpha(\pr^{-1}(\pr(\Game)))$, i.e., all vertices attracted
to the current highest vertices of priority $p := \pr(\Game)$ for player $\alpha=p \bmod 2$,
and recursively computes the winning regions $(W_\Even,W_\Odd)$ of the remaining subgame $\Game\setminus A$.
If the opponent $\invalpha$ can attract vertices in $A$ to $W_\invalpha$, then $\invalpha$ wins $W'_\invalpha := \text{Attr}^\Game_\invalpha(W_\invalpha)$ and the solution for the remainder $\Game\setminus W'_\invalpha$ is computed recursively.
Otherwise, $\alpha$ wins $A$ and no further recursion is necessary.
The strategies for both players are trivially obtained during attractor computation and by assigning to winning $p$-vertices in $A$ any strategy to vertices in $W_\alpha\cup A$.

Zielonka's original algorithm has been extended and improved over the years.
In his thesis, Verver~\cite{Verver2013} improves the partitioning of the game
after computing $A$ by extending $A$ with the attractors of the next highest vertices if they are of the same parity.
The original algorithm always recomputes the solution of $\Game\setminus W'_\invalpha$ if $W_\invalpha$ is nonempty, even if no vertices are attracted to $W_\invalpha$.
Liu et al. propose that this is not necessary~\cite{DBLP:conf/tase/LiuDT14}.
See Alg.~\ref{alg:zielonka} for the recursive algorithm with these modifications.
Other extensions that we do not consider here are the subexponential algorithm~\cite{DBLP:journals/siamcomp/JurdzinskiPZ08} and the big steps algorithm~\cite{DBLP:journals/jcss/Schewe17} that have been reported to perform slower
than ordinary Zielonka~\cite{DBLP:conf/atva/FriedmannL09}.
Also, variations using BDDs have been proposed~\cite{DBLP:conf/mochart/BakeraEKR08,DBLP:journals/corr/KantP14}.

Although the implementation of the recursive algorithm in PGsolver~\cite{DBLP:conf/atva/FriedmannL09} is typically used for comparisons in the literature,
improved implementations have been proposed by Verver~\cite{Verver2013}, Di Stasio et al.~\cite{DBLP:conf/facs2/StasioMPS14}, Liu et al.~\cite{DBLP:conf/tase/LiuDT14}, and Arcucci et al.~\cite{DBLP:conf/iccS/ArcucciMMS17}.
Verver suggests to record the number of remaining ``escaping'' edges for each vertex during attractor computation, to reduce the complexity of attractor computation at the cost of an extra integer per vertex.
Di Stasio et al. avoid creating copies of the game for recursive operations by recording which vertices are removed in a special array.
Recently, Arcucci et al. extended the implementation in~\cite{DBLP:conf/facs2/StasioMPS14} with a multi-core implementation of attractor computation~\cite{DBLP:conf/iccS/ArcucciMMS17}.

\begin{algorithm}[t]
	\Def{\Zielonka{$\Game$}}{
		\lIf{$\Game = \emptyset$}{\Return $\emptyset, \emptyset$}
		$\alpha$ $\leftarrow$ $\pr(\Game)$ mod 2 \;
		$A$ $\leftarrow$ \ZielonkaAttr{$\Game$, $\alpha$} \;
		$W_\Even, W_\Odd$ $\leftarrow$ \Zielonka{$\Game\setminus A$} \;
		$W'_{\invalpha}$ $\leftarrow$ $\text{Attr}^{\Game}_{\invalpha}(W_{\invalpha})$ \;
		\If{$W'_{\invalpha} = W_{\invalpha}$}{
			$W_{\alpha}$ $\leftarrow$ $W_\alpha \cup A$
		}
		\Else{
			$W_\Even, W_\Odd$ $\leftarrow$ \Zielonka{$\Game\setminus W'_{\invalpha}$} \;
			$W_{\invalpha}$ $\leftarrow$ $W_{\invalpha} \cup W'_{\invalpha}$
		}
		\Return {$W_{\Even}, W_{\Odd}$} \;
	}
	\BlankLine
	\Def{\ZielonkaAttr{$\Game$, $\alpha$}}{
		$A$ $\leftarrow$ $\emptyset$ \;
		\lWhile{$\pr(\Game \setminus A) =_2 \alpha$}{
			$A$ $\leftarrow$ $A \cup \text{Attr}^{\Game\setminus A}_\alpha(\pr^{-1}(\pr(\Game \setminus A))$
		}
		\Return $A$
	}
	\caption{The recursive Zielonka algorithm.}
	\label{alg:zielonka}
\end{algorithm}

The implementation in Oink is based upon the ideas described above.
Furthermore, we improve the implementation using the following techniques.
\begin{itemize}
	\item Instead of creating copies of the ``removed'' array (\cite{DBLP:conf/facs2/StasioMPS14}) for each recursive step, we use a single ``region'' array that stores for each vertex that it is attracted by the $r$th call to \texttt{attr}.
	This value is initially $\bot$ for all vertices and is reset to $\bot$ for vertices in  $\Game\setminus W'_\invalpha$ (line~10).
	We record the initial $r$ at each depth and thus derive that all vertices with a value $\geq r$ or $\bot$ are part of the subgame.
	\item As a preprocessing step, we order all vertices by priority. We can then quickly obtain the highest vertex of each subgame.
	\item We eliminate the recursion using a stack.
	\item We implement an alternative lock-free attractor, relying on
	the work-stealing library Lace~\cite{DBLP:conf/europar/DijkP14} that provides
	fine-grained load balancing.
\end{itemize}

In the interest of space, we cannot describe the multi-core attractor in-depth.
This implementation is fairly straightforward.
We implement the attractor recursively where the work-stealing framework runs the recursive operations in parallel.
Like typical lock-free algorithms, we rely on the \texttt{compare-and-swap} operation to implement safe communication between threads.
The attractor uses this operation when manipulating the number of escaping edges and to ``claim'' a vertex by setting its value in the region array from $\bot$ to $r$.

\subsubsection{Empirical Evaluation}

\begin{table}[t]
	\begin{tabu} to \linewidth {X[2]X[2r]X[c]X[2r]X[c]X[2r]X[c]X[2r]X[c]}
	\toprule
	& \multicolumn{2}{c}{Model checking} & \multicolumn{2}{c}{Equiv checking} & \multicolumn{2}{c}{Random games} & \multicolumn{2}{c}{Total} \\
	\midrule
	\input{R/zlk}
	\bottomrule
\end{tabu}
\vspace{.5em}
\caption{Runtimes in sec. (PAR2) and number of timeouts (15 minutes) of the four solvers PGSolver (\texttt{pgzlk}), SPGSolver (\texttt{spg}), pbespgsolve (\texttt{pbeszlk}) and Oink (sequential \texttt{zlk}, multi-core \texttt{zlk-1} and \texttt{zlk-8}, unoptimized \texttt{uzlk}).}
	\label{tbl:zlk}
\end{table}

We compare our implementation of Zielonka's recursive algorithm with and without the optimizations of Alg.~\ref{alg:zielonka} to PGSolver, to Verver's implementation pbespgsolve~\cite{DBLP:conf/tacas/CranenGKSVWW13,Verver2013} and to SPGSolver~\cite{DBLP:conf/iccS/ArcucciMMS17,DBLP:conf/facs2/StasioMPS14}. Unfortunately, the Java version of SPGSolver (all three variations) suffers from severe performance degradation for unknown reasons. They also provide a C++ implementation in their online repository, which we used instead. The multi-core version of the SPGSolver tool relies on \texttt{async} tasks provided by C++11.
Similar to the previous sections, we disable the optional preprocessors that solve single parity games, remove self-loops and solve winner-controlled winning cycles.

See Table~\ref{tbl:zlk}.
The results show that the implementation in Oink outperforms PGSolver by several orders of magnitude on all benchmark types.
PGSolver timed out for $83$ of all $949$ games.
%The implementation in Oink is competitive with the solvers \texttt{spg-seq} and \texttt{pbeszlk}.
%Solver \texttt{zlk} is competitive with the solvers \texttt{spg-seq}, considering that \texttt{spg-seq} timed out for $3$ games.
The solvers \texttt{spg-seq} and \texttt{pbeszlk} are faster than Oink on the model checking and equivalence checking games, but are significantly outperformed on random games.
We also observe severe performance degradation for \texttt{spg-mc} on random games.
It appears that our parallel implementation of Zielonka's algorithm also does not scale well.
Finally, there seems to be no significant difference between the optimized and unoptimized versions of Zielonka's algorithm, except for random games.
%A surprising result is that \texttt{zlk-1} is faster overall than the sequential \texttt{zlk} solver.
%This may be due to factors such as different memory access patterns, as \texttt{zlk} uses a queue to compute attractors whereas \texttt{zlk-1} uses task-based recursion.
%Finally, we observe that the implementation in Oink has similar performance as the implementation in pbespgsolve.

%% file: R/zlk.tex
% latex table generated in R 3.4.3 by xtable 1.8-2 package
% 
  zlk-8 & 94 & 0 & 415 & 0 & 11 & 0 & \textbf{521} & 0 \\ 
  zlk & 88 & 0 & 472 & 0 & \textbf{6} & 0 & 566 & 0 \\ 
  zlk-1 & 97 & 0 & 512 & 0 & 7 & 0 & 616 & 0 \\ 
  uzlk & 89 & 0 & 472 & 0 & 69 & 0 & 630 & 0 \\ 
  pbeszlk & 64 & 0 & 513 & 0 & 338 & 0 & 915 & 0 \\ 
  spg-seq & \textbf{58} & 0 & \textbf{198} & 0 & 694 & 0 & 950 & 0 \\ 
  spg-mc & 389 & 0 & 1451 & 0 & 72608 & 37 & 74447 & 37 \\ 
  pgzlk & 65905 & 33 & 68013 & 36 & 41629 & 14 & 175547 & 83 \\ 
  

%% file: prioprom.tex
%!TEX root = main.tex

In recent work, a new family of algorithms has been proposed based on \emph{priority promotion}~\cite{DBLP:conf/cav/BenerecettiDM16}.
Priority promotion starts with a similar decomposition of the game as Zielonka's recursive
algorithm.
Priority promotion is based on the insight that a recursive decomposition based on attractor computation leads to regions with a specific property related to the highest priority in the region, called its \emph{measure} $p$.
This property is that all plays that stay in the region are won by the player who wins the highest priority $p$, denoted by player $\alpha$.
The other player $\invalpha$ has three options. They either lose the game by staying in the region, or they can leave the region by playing to an $\alpha$-region of higher measure, or they can leave the region to a lower region of either player via a vertex with priority $p$. 
The goal of $\alpha$ is to find ``closed'' $\alpha$-regions, where $\invalpha$ cannot escape to lower regions.
The result is a region where player $\invalpha$ either loses, or leaves the region to a higher $\alpha$-region which may or may not be closed.
The measure of the closed $\alpha$-region is then ``promoted'' to the measure of the lowest higher region to which $\invalpha$ can escape and the attractor-based decomposition is recomputed for all lower regions.
The promoted region may now attract from regions with a measure between its original measure and its promoted measure, thus requiring recomputing the decomposition.
When player $\invalpha$ cannot escape from an $\alpha$-region to a higher $\alpha$-region, player $\alpha$ is the winner of all vertices in the region.

Priority promotion was proposed in \cite{DBLP:conf/cav/BenerecettiDM16} and improved in~\cite{DBLP:journals/corr/BenerecettiDM16,DBLP:conf/hvc/BenerecettiDM16}.
The original PP algorithm~\cite{DBLP:conf/cav/BenerecettiDM16} forgets all progress (``resets'') in lower regions after promotion.
The PP+ algorithm~\cite{DBLP:journals/corr/BenerecettiDM16} only resets lower regions of player $\invalpha$.
The RR algorithm~\cite{DBLP:conf/hvc/BenerecettiDM16} only resets \emph{some} lower regions of player $\invalpha$.
The DP algorithm~\cite{DBLP:journals/corr/BenerecettiDM16} uses a heuristic to delay certain promotions to avoid resets.
We implement all four algorithms and also combine the DP algorithm, which is based on PP+, with the RR algorithm.

\subsubsection{Empirical Evaluation}

\begin{table}[t]
	\begin{tabu} to \linewidth {X[2]X[2r]X[c]X[2r]X[c]X[2r]X[c]X[2r]X[c]}
		\toprule
		& \multicolumn{2}{c}{Model checking} & \multicolumn{2}{c}{Equiv checking} & \multicolumn{2}{c}{Random games} & \multicolumn{2}{c}{Total} \\
		\midrule
		\input{R/pp}
		\bottomrule
	\end{tabu}
	\vspace{.5em}
	\caption{Runtimes in sec. (PAR2) and number of timeouts (15 minutes) of the five priority promotion solvers in Oink.}
\label{tbl:pp}
\end{table}

We compare our implementation of five variations of priority promotion in Oink.
As we do not compare with other solvers, we enable the optional preprocessors that solve single parity games, remove self-loops and solve winner-controlled winning cycles.

See Table~\ref{tbl:pp}.
Overall, we see that the simplest solver \texttt{pp} performs just as good as the more complex solvers. %, but only slightly better than the other solvers.
The motivation for the variations is based on crafted families that require an exponential number of promotions.
The \texttt{pp} solver may be most vulnerable to these crafted families, but on practical games and random games there is no significant difference.

%% file: R/pp.tex
% latex table generated in R 3.4.3 by xtable 1.8-2 package
% 
  ppp & \textbf{81} & 0 & \textbf{382} & 0 & \textbf{12} & 0 & \textbf{475} & 0 \\ 
  pp & 82 & 0 & \textbf{382} & 0 & \textbf{12} & 0 & 476 & 0 \\ 
  rr & \textbf{81} & 0 & 385 & 0 & \textbf{12} & 0 & 477 & 0 \\ 
  dp & 84 & 0 & 389 & 0 & 15 & 0 & 488 & 0 \\ 
  rrdp & 83 & 0 & 394 & 0 & 14 & 0 & 491 & 0 \\ 
  

%% file: conclusions.tex
%!TEX root = main.tex

\begin{table}[t]
	\begin{tabu} to \linewidth {X[2]X[2r]X[c]X[2r]X[c]X[2r]X[c]X[2r]X[c]}
		\toprule
		& \multicolumn{2}{c}{Model checking} & \multicolumn{2}{c}{Equiv checking} & \multicolumn{2}{c}{Random games} & \multicolumn{2}{c}{Total} \\
		\midrule
		\input{R/oink}

		\bottomrule
	\end{tabu}
	\vspace{.5em}
	\caption{Runtimes in sec. (PAR2) and number of timeouts (15 minutes) of the sequential implementations of the five solvers in Oink described in this paper.}
	\label{tbl:oink}
\end{table}

\begin{figure}[t]
	\resizebox{\linewidth}{!}{
		\input{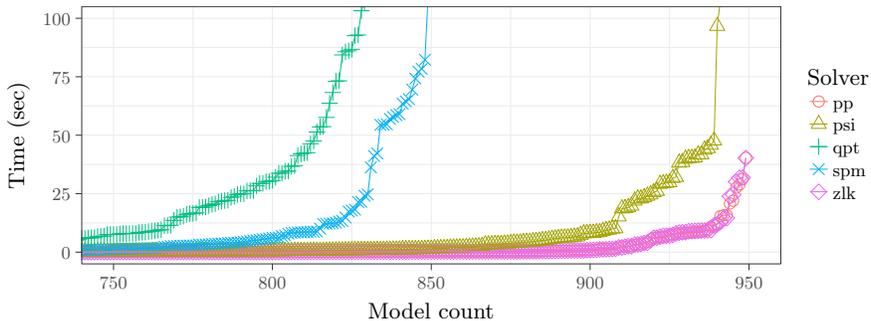}
	}
	\caption{A cactus plot of five sequential solvers implemented in Oink. The plot shows how many games are (individually) solved within a certain amount of time.}
	\label{fig:oink_in}
\end{figure}

See Table~\ref{tbl:oink} for a comparison of the five main sequential algorithms in Oink, including the preprocessing that removes winner-controlled winning cycles, self-loops and solves single parity games.
The results show that the \texttt{zlk} and the \texttt{pp} solvers have similar performance and outperform the other solvers.
See also Fig.~\ref{fig:oink_in} for a {cactus plot} of these five solvers.

Priority promotion is a powerful and attractive idea,
as promoting closed $\alpha$-regions is similar to \emph{cap-and-carryover} in small progress measures.
Attractor computation finds such regions directly whereas value iteration algorithms may require many iterations.
We confirm the observations in~\cite{DBLP:conf/cav/BenerecettiDM16} that the algorithm has a good performance but it is not faster than Zielonka's algorithm.

In this work, we studied modern parity game algorithms using a new tool named \textbf{Oink}.
Oink is publicly available via \url{https://www.github.com/trolando/oink}.
We implemented a number of modern algorithms and provided a comprehensive description of these algorithms, introducing \emph{cap-and-carryover} to understand small progress measures.
We proposed improvements to strategy improvement and to Zielonka's algorithm.
We presented an empirical evaluation of Oink, comparing its performance with state-of-the-art solvers, especially the popular PGSolver tool.
The results demonstrate that Oink is competitive with other implementations and in fact outperforms PGSolver for all algorithms, especially Zielonka's recursive algorithm.
This result is particularly interesting considering that many publications compare the performance of novel ideas to Zielonka's algorithm in PGSolver.

\section*{Acknowledgements}

We thank Tim Willemse and John Fearnley for their helpful comments
and Jaco van de Pol for the use of their computer cluster.

%% file: R/oink.tex
% latex table generated in R 3.4.3 by xtable 1.8-2 package
% 
  pp & 82 & 0 & \textbf{382} & 0 & 12 & 0 & \textbf{476} & 0 \\ 
  zlk & \textbf{78} & 0 & 393 & 0 & \textbf{10} & 0 & 481 & 0 \\ 
  psi & 231 & 0 & 2440 & 0 & 689 & 0 & 3359 & 0 \\ 
  spm & 1007 & 0 & 3079 & 0 & 156885 & 87 & 160971 & 87 \\ 
  qpt & 59559 & 31 & 60728 & 31 & 62104 & 33 & 182391 & 95 \\

%% file: main.bbl
\begin{thebibliography}{10}
\providecommand{\url}[1]{\texttt{#1}}
\providecommand{\urlprefix}{URL }

\bibitem{DBLP:conf/iccS/ArcucciMMS17}
Arcucci, R., Marotta, U., Murano, A., Sorrentino, L.: Parallel parity games: a
  multicore attractor for the {Zielonka} recursive algorithm. In: {ICCS}.
  Procedia Computer Science, vol. 108, pp. 525--534. Elsevier (2017)

\bibitem{DBLP:conf/mochart/BakeraEKR08}
Bakera, M., Edelkamp, S., Kissmann, P., Renner, C.D.: Solving mu-calculus
  parity games by symbolic planning. In: MoChArt. LNCS, vol. 5348, pp. 15--33.
  Springer (2008)

\bibitem{DBLP:journals/corr/BenerecettiDM16}
Benerecetti, M., Dell'Erba, D., Mogavero, F.: {A Delayed Promotion Policy for
  Parity Games}. In: {GandALF} 2016. {EPTCS}, vol. 226, pp. 30--45 (2016)

\bibitem{DBLP:conf/hvc/BenerecettiDM16}
Benerecetti, M., Dell'Erba, D., Mogavero, F.: {Improving Priority Promotion for
  Parity Games}. In: {HVC} 2016. LNCS, vol. 10028, pp. 117--133 (2016)

\bibitem{DBLP:conf/cav/BenerecettiDM16}
Benerecetti, M., Dell'Erba, D., Mogavero, F.: {Solving Parity Games via
  Priority Promotion}. In: {CAV} 2016. LNCS, vol. 9780, pp. 270--290. Springer
  (2016)

\bibitem{vanderBerg2010}
van~der Berg, F.: Solving parity games on the playstation 3. In: Twente Student
  Conference (2010)

\bibitem{DBLP:journals/dam/BjorklundV07}
Bj{\"{o}}rklund, H., Vorobyov, S.G.: A combinatorial strongly subexponential
  strategy improvement algorithm for mean payoff games. Discrete Applied
  Mathematics  155(2),  210--229 (2007)

\bibitem{BootsmaW13}
Bootsma, P.: Speeding up the small progress measures algorithm for parity games
  using the GPU. Master's thesis, Eindhoven University of Technology (2013)

\bibitem{DBLP:conf/stoc/CaludeJKL017}
Calude, C.S., Jain, S., Khoussainov, B., Li, W., Stephan, F.: Deciding parity
  games in quasipolynomial time. In: {STOC}. pp. 252--263. {ACM} (2017)

\bibitem{DBLP:conf/csl/ChatterjeeDHL17}
Chatterjee, K., Dvor{\'{a}}k, W., Henzinger, M., Loitzenbauer, V.: Improved
  set-based symbolic algorithms for parity games. In: {CSL}. LIPIcs, vol.~82,
  pp. 18:1--18:21. Schloss Dagstuhl - Leibniz-Zentrum fuer Informatik (2017)

\bibitem{DBLP:conf/tacas/CranenGKSVWW13}
Cranen, S., Groote, J.F., Keiren, J.J.A., Stappers, F.P.M., de~Vink, E.P.,
  Wesselink, W., Willemse, T.A.C.: An overview of the mcrl2 toolset and its
  recent advances. In: {TACAS}. LNCS, vol. 7795, pp. 199--213. Springer (2013)

\bibitem{DBLP:conf/europar/DijkP14}
van Dijk, T., van~de Pol, J.C.: Lace: Non-blocking split deque for
  work-stealing. In: Euro-Par Workshops {(2)}. LNCS, vol. 8806, pp. 206--217.
  Springer (2014)

\bibitem{DBLP:conf/focs/EmersonJ91}
Emerson, E.A., Jutla, C.S.: Tree automata, mu-calculus and determinacy
  (extended abstract). In: {FOCS}. pp. 368--377. {IEEE} Computer Society (1991)

\bibitem{DBLP:conf/cav/EmersonJS93}
Emerson, E.A., Jutla, C.S., Sistla, A.P.: On model-checking for fragments of
  mu-calculus. In: {CAV}. LNCS, vol. 697, pp. 385--396. Springer (1993)

\bibitem{DBLP:journals/tcs/EmersonJS01}
Emerson, E.A., Jutla, C.S., Sistla, A.P.: On model checking for the mu-calculus
  and its fragments. Theor. Comput. Sci.  258(1-2),  491--522 (2001)

\bibitem{DBLP:conf/lpar/Fearnley10}
Fearnley, J.: Non-oblivious strategy improvement. In: {LPAR} (Dakar). LNCS,
  vol. 6355, pp. 212--230. Springer (2010)

\bibitem{DBLP:conf/cav/Fearnley17}
Fearnley, J.: Efficient parallel strategy improvement for parity games. In:
  {CAV} {(2)}. LNCS, vol. 10427, pp. 137--154. Springer (2017)

\bibitem{DBLP:conf/spin/FearnleyJS0W17}
Fearnley, J., Jain, S., Schewe, S., Stephan, F., Wojtczak, D.: An ordered
  approach to solving parity games in quasi polynomial time and quasi linear
  space. In: {SPIN}. pp. 112--121. {ACM} (2017)

\bibitem{DBLP:conf/atva/FriedmannL09}
Friedmann, O., Lange, M.: Solving parity games in practice. In: {ATVA}. {LNCS},
  vol. 5799, pp. 182--196. Springer (2009)

\bibitem{DBLP:conf/gandalf/Friedmann2010}
Friedmann, O., Lange, M.: Local strategy improvement for parity game solving.
  In: {GANDALF}. {EPTCS}, vol.~25, pp. 118--131 (2010)

\bibitem{DBLP:journals/corr/GazdaW13}
Gazda, M., Willemse, T.A.C.: Zielonka's recursive algorithm: dull, weak and
  solitaire games and tighter bounds. In: GandALF. {EPTCS}, vol. 119, pp. 7--20
  (2013)

\bibitem{DBLP:journals/corr/GazdaW15}
Gazda, M., Willemse, T.A.C.: Improvement in small progress measures. In:
  GandALF. {EPTCS}, vol. 193, pp. 158--171 (2015)

\bibitem{DBLP:conf/dagstuhl/2001automata}
Gr{\"{a}}del, E., Thomas, W., Wilke, T. (eds.): Automata, Logics, and Infinite
  Games: {A} Guide to Current Research [outcome of a Dagstuhl seminar, February
  2001], LNCS, vol. 2500. Springer (2002)

\bibitem{DBLP:conf/atva/HoffmannL13}
Hoffmann, P., Luttenberger, M.: Solving parity games on the {GPU}. In: {ATVA}.
  LNCS, vol. 8172, pp. 455--459. Springer (2013)

\bibitem{DBLP:conf/hvc/HuthKP11}
Huth, M., Kuo, J.H., Piterman, N.: Concurrent small progress measures. In:
  Haifa Verification Conference. LNCS, vol. 7261, pp. 130--144. Springer (2011)

\bibitem{DBLP:journals/ipl/Jurdzinski98}
Jurdzinski, M.: Deciding the winner in parity games is in {UP} $\cap$ co-{UP}.
  Inf. Process. Lett.  68(3),  119--124 (1998)

\bibitem{DBLP:conf/stacs/Jurdzinski00}
Jurdzinski, M.: Small progress measures for solving parity games. In: {STACS}.
  LNCS, vol. 1770, pp. 290--301. Springer (2000)

\bibitem{DBLP:conf/lics/JurdzinskiL17}
Jurdzinski, M., Lazic, R.: Succinct progress measures for solving parity games.
  In: {LICS}. pp. 1--9. {IEEE} Computer Society (2017)

\bibitem{DBLP:journals/siamcomp/JurdzinskiPZ08}
Jurdzinski, M., Paterson, M., Zwick, U.: A deterministic subexponential
  algorithm for solving parity games. {SIAM} J. Comput.  38(4),  1519--1532
  (2008)

\bibitem{DBLP:journals/corr/KantP14}
Kant, G., van~de Pol, J.: Generating and solving symbolic parity games. In:
  {GRAPHITE}. {EPTCS}, vol. 159, pp. 2--14 (2014)

\bibitem{DBLP:conf/fsen/Keiren15}
Keiren, J.J.A.: Benchmarks for parity games. In: {FSEN}. LNCS, vol. 9392, pp.
  127--142. Springer (2015)

\bibitem{DBLP:journals/tcs/Kozen83}
Kozen, D.: Results on the propositional mu-calculus. Theor. Comput. Sci.  27,
  333--354 (1983)

\bibitem{DBLP:conf/stoc/KupfermanV98}
Kupferman, O., Vardi, M.Y.: Weak alternating automata and tree automata
  emptiness. In: {STOC}. pp. 224--233. {ACM} (1998)

\bibitem{DBLP:conf/tase/LiuDT14}
Liu, Y., Duan, Z., Tian, C.: An improved recursive algorithm for parity games.
  In: {TASE}. pp. 154--161. {IEEE} Computer Society (2014)

\bibitem{DBLP:journals/corr/abs-0806-2923}
Luttenberger, M.: Strategy iteration using non-deterministic strategies for
  solving parity games. CoRR  abs/0806.2923 (2008)

\bibitem{DBLP:conf/atva/MeyerL16}
Meyer, P.J., Luttenberger, M.: Solving mean-payoff games on the {GPU}. In:
  {ATVA}. LNCS, vol. 9938, pp. 262--267 (2016)

\bibitem{DBLP:journals/entcs/PolW08}
van~de Pol, J., Weber, M.: A multi-core solver for parity games. Electr. Notes
  Theor. Comput. Sci.  220(2),  19--34 (2008)

\bibitem{DBLP:conf/csl/Schewe08}
Schewe, S.: An optimal strategy improvement algorithm for solving parity and
  payoff games. In: {CSL}. LNCS, vol. 5213, pp. 369--384. Springer (2008)

\bibitem{DBLP:journals/jcss/Schewe17}
Schewe, S.: Solving parity games in big steps. J. Comput. Syst. Sci.  84,
  243--262 (2017)

\bibitem{DBLP:conf/facs2/StasioMPS14}
Stasio, A.D., Murano, A., Prignano, V., Sorrentino, L.: Solving parity games in
  scala. In: {FACS}. LNCS, vol. 8997, pp. 145--161. Springer (2014)

\bibitem{Verver2013}
Verver, M.: Practical improvements to parity game solving. Master's thesis,
  University of Twente (2013)

\bibitem{DBLP:conf/cav/VogeJ00}
V{\"{o}}ge, J., Jurdzinski, M.: A discrete strategy improvement algorithm for
  solving parity games. In: {CAV}. LNCS, vol. 1855, pp. 202--215. Springer
  (2000)

\bibitem{DBLP:journals/tcs/Zielonka98}
Zielonka, W.: Infinite games on finitely coloured graphs with applications to
  automata on infinite trees. Theor. Comput. Sci.  200(1-2),  135--183 (1998)

\end{thebibliography}
